Concept Paper:

# Introducing the Unitychain Structure

*A novel blockchain-like structure that enables greater parallel processing, security, and performance for networks that leverage distributed key generation and classical consensus protocols*

December 31st, 2020

Joshua D. Tobkin

Almost all "Proof of Stake" blockchains that leverage threshold cryptography for *unique* randomness to run any variety of classical consensus algorithms require some form of distributed key generation (DKG) ritual [1], and thus a *prolonged down-time* for nodes to reshuffle and configure their membership [2].

"Unitychain" is a novel blockchain-*like* structure that drastically improves transaction scalability and security, while maintaining *ongoing* network performance, even if participating nodes are required to perform a new Distributed Key Generation (DKG) procedure for security purposes.

The Unitychain structure, furthermore, enables greater parallel processing by the assignment of different network node configurations for various database and compute ranges *into multiple strands of blockchains that intersect*, creating a multi-helix structure, which we call a "Unitychain." This thereby enables the network to further bifurcate the roles of nodes into arbitrary yet deterministic network responsibilities in order to maximize the global compute potential.

During the intersecting strands, the network's responsibility will be assigned to one of the network's node configurations, which takes over the entire database key-range and compute responsibilities for the cycle, while the other network configuration reshuffles its membership into a new logical arrangement. This generates a deterministic, yet unpredictable and unique *network topology*, thereby increasing the overall network randomness, which is required for sustained security [3].

Whereas most blockchain designs that leverage distributed key generation as part of their protocol would otherwise require a prolonged downtime duration in order to reconfigure node membership [4], thus affecting Liveness guarantees, a Unitychain is designed to enable ongoing operations while participating nodes run sub-routines to reshuffle their other network configurations held in the intersecting strands of the structure.

This creates the experience for the end-user transacting on the network that the network never experiences any downtime. This is because as one strand of the blockchained network configuration is reshuffling its constitution, the node configuration of the other(s) resume the responsibilities of the entire key-space or logical partition for its counterpart strand.

## The Main Benefits of the Unitychain Structure:

The Unitychain structure removes the need for the network to require downtime affecting the network's liveness during reshuffling node membership and generating new distributed key pairs, which is otherwise required in order to leverage certain powerful cryptographic primitives, like BLS threshold signatures, whose security properties diminish over time [5] — since we assume nodes can and may *eventually* collude given the opportunity. Therefore, protocols that leverage unique threshold signatures that require a distributed key generation procedure must randomly reshuffle node membership *often* to prevent malicious and collusive behaviour from more easily forming.

Pedantically, the Unitychain structure enables blockchain protocols that rely on cryptographic primitives, *which require DKG node rituals for setup*, the ability to frequently reshuffle their node membership to generate *newly*-formed DKG pairings, while continuously ensuring maximum security and uninterrupted performance for the operating network [6]. The Unitychain's scalability advantages are achieved by apexing each node's potential compute capacity by allowing each singular node to hold multiple logical positions in the network *concurrently*. This results in overall improved load-balancing and substantially greater parallel processing capabilities for the network, while maintaining and ensuring the continued safety and security of the structure.

## Nomenclature:

*Description and definitions of Components*

**Byzantine Fault Tolerance (BFT):** Byzantine Fault Tolerance is the characteristic which defines a system that tolerates the class of failures that belong to the Byzantine Generals' Problem.

**Unitychain:** A "Unitychain" is a multi-helix blockchain structure with 2 or more intersecting blockchain strands whose blocks contain different network configurations of node membership, blockchain transactions, public key data, or arbitrary state.

**Epochs:** Period of time of arbitrary length measured in a number of Cycles, during which nodes can join and exit the blockchain to provide consensus and storage services.

**Epoch Blocks:** Epoch blocks are data structures which are signed by nodes every Epoch which contains vital data such as the list of nodes allowed to participate in the network during the next period, a summary or reference to the transactions that were processed in the prior block, as well as other data required to drive the consensus protocol forward.

**Cycles:** Cycles are a shorter period of time, for example 10 seconds or 5 minutes (tunable parameter), which comprise an Epoch.

**Cycle Blocks:** Cycle blocks are the blocks that are produced and majority signed to represent the start and end of a Cycle. Cycle blocks contain a summary or a reference to the transactions that were processed for the Cycle and the nodes who participated in consensus.

**Ascending Strand:** The blockchain strand that is ascending takes dominance during Epoch Blocks.

**Descending Strand:** The blockchain strand that is descending takes a submissive role during Epoch Blocks. In the prior Epoch, this strand was the dominant strand.

**Dominant Strand:** The blockchain strand that takes over the responsibilities of submissive strand during Epoch Blocks.

**Submissive Strand:** The blockchain strand that relinquishes the responsibilities during Epoch Blocks and which subsequently reconstitutes its node membership through reshuffling.

**Parallel Universes:** Nodes in our system can play more than 1 logical position at a time, since their node identity information and randomized position is held in multiple strands of the Unitychain structure.

**Key Range Partitioning:** Partitioning data based on ranges of values in the database, for example, into quadrants, or even positive and negative key-values, or other identifiers.

**Distributed Key Generation:** Distributed key generation (DKG) is a cryptographic process in which multiple parties contribute to the calculation of a shared public and private key set.

**Verifiable Random Function (VRF):** A Verifiable Random Function (VRF) is a cryptographic primitive that maps inputs to verifiable pseudorandom outputs.

**Origin Block:** At the beginning of a Unitychain is a singular "Origin Block," which contains the node membership of all nodes as well as any other relevant data required for the protocol to initiate.

**Threshold Signatures:** Threshold signatures are a type of cryptographic signature in which only n out of m are required to sign a message together to create a valid group signature. Any combination of n out m signatures are required to achieve a valid group signature.

## How Does the Unitychain Structure Function?

Most Blockchains that run classical consensus algorithms that have leader elections from committees of nodes must know the network membership in order to select the next leader or a cohort of nodes [7]. Node identities are commonly therefore registered in a fixed ordered list inside a "block" of a blockchain and is used as a collectively agreed upon source of truth for active nodes to sample from for leader-election or other core roles to drive the protocol forward.

A "Unitychain" is a multi-helix *blockchain-like* structure with 2 or more intersecting blockchain strands whose blocks contain different network configurations of node membership, blockchain transaction data or arbitrary state. This structure enables the dividing-up of network responsibilities, for example by key-ranges of the database, database shards, or any arbitrary logical partition, enabling more scalability and security without any network downtime.

Nodes may exist in multiple "Parallel Universes," or in *all strands at the same time*, holding different logical network positions and topologies in each strand. This data is known to all nodes that follow protocol and is held within each block of each strand in the Unitychain structure. All active nodes are in possession of all blocks of all Unitychain strands and are, therefore, able to trivially compute if the correct nodes involved in consensus are the correct ones based on the protocol rules and the data *stored in the last verified and majority signed block of the strand.*

In one Unitychain manifestation, of many possible innovations that embrace the spirit of this design, suppose all *even* key-ranges of the database are the responsibility of the "Positive Strand" and all the *odd* key-ranges are the responsibility of the "Negative Strand." **See Figure #1.**

**Figure #1**

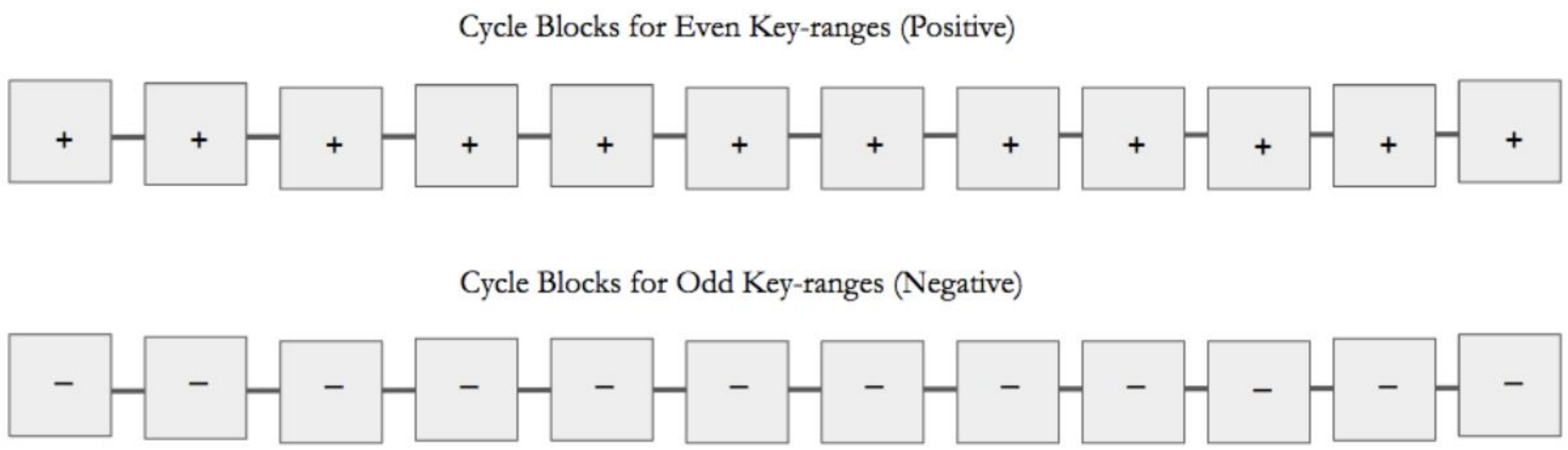

Each Cycle a new block in each strand is proposed and signed by the majority of nodes of the prior block configuration. Cycle blocks contain a record or reference of all the transactions that were processed during the Cycle, among other network reference data including, but not limited to, node public key data and a node's *logical position* in the network.

A majority aggregate threshold signature is verifiable based on the knowledge of the network *topology of the prior block for each strand.* Computing network consensus by verifying this final aggregated cryptographic signature on all core protocol decisions is trivial. This enables even commodity hardware like modern-day laptops to rapidly become suitable to participate in consensus and network services.

For example, suppose each Cycle block is approximately 10 seconds in duration (tunable system parameter). This means there would be about 6 Cycles every minute (another tunable parameter). Each Cycle block is majority signed by the node configuration of the prior block in the strand. Most importantly, each Cycle can present a newly formed and majority agreed upon network configuration, effectively enabling the entire network to frequently morph into a new random topology, thereby *improving overall system security*. Moreover, by allowing nodes to exist in multiple strands, we can better maximize a node's potential latent compute resources providing overall increased scalability and throughput on an individual node basis. In the following example, we call the 1st and 7th block created thereafter "Epoch Blocks," which are the blocks wherein multiple strands of the Unitychain intersect. **See Figure #2**

**Figure #2**

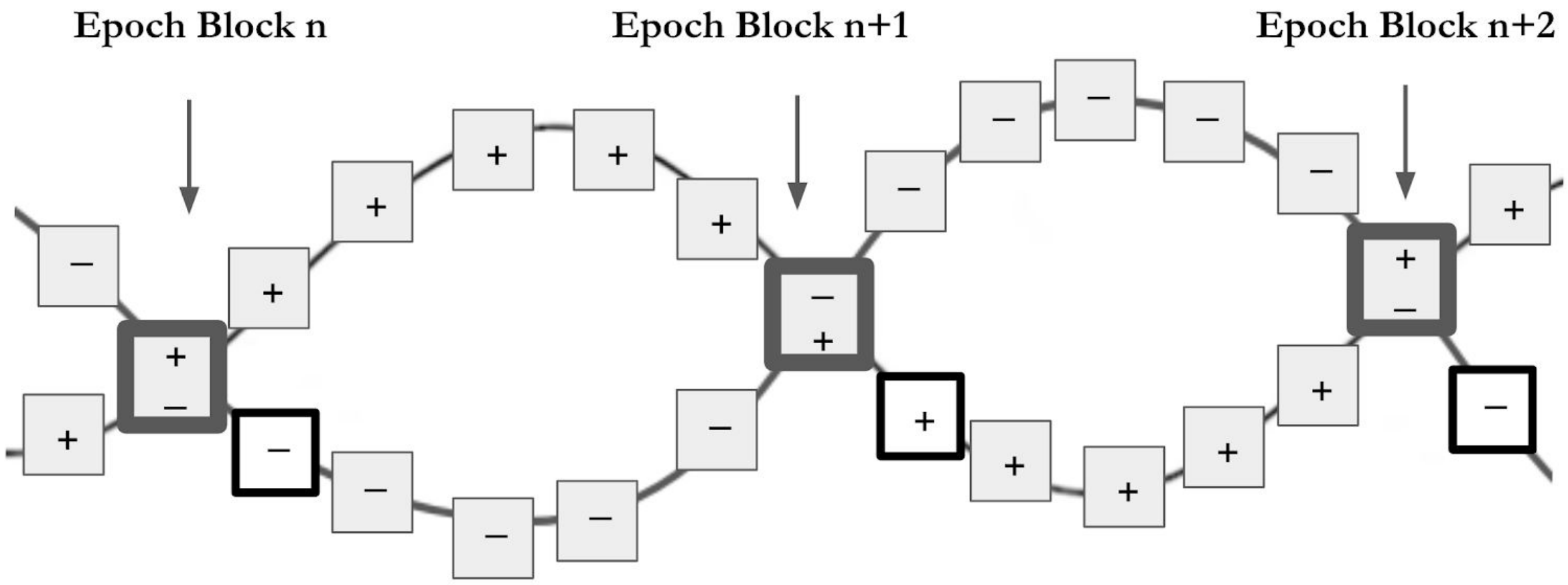

Epoch Blocks are cryptographically signed multiple times by the majority of nodes in all the intersecting strands. During Epoch Blocks, nodes are agreeing that the node configuration of the *Ascending Strand* deterministically shall take over the entire database responsibility of the partitioned key-range or other logical segregation of roles that the prior dominant strand had otherwise covered *(see above for nomenclature)*.

For example, suppose one strand of the Unitychain is covering all *odd* (-) keys-ranges, while the other strand covers all *even* (+) keys.

**Depicted in Figure #3 below**, during "Epoch Block n," the Ascending Strand has a *negative (odd) valence* and therefore the node configuration of this strand will inherit the entire responsibility of the network during this Epoch Block, while the nodes in the descending strand begins its reshuffling procedures.

Whereas, in "Epoch Block n+1" the Ascending Strand has a *positive (+) valence* whose node configuration will take up the collective responsibilities for the current Epoch, while the other strand reshuffles its membership allowing new nodes to join and leave the network. This continues so on and so forth *into perpetuity* allowing the verifiably-morphing node composition of the network to remain hyper-fluid.

**Figure #3**

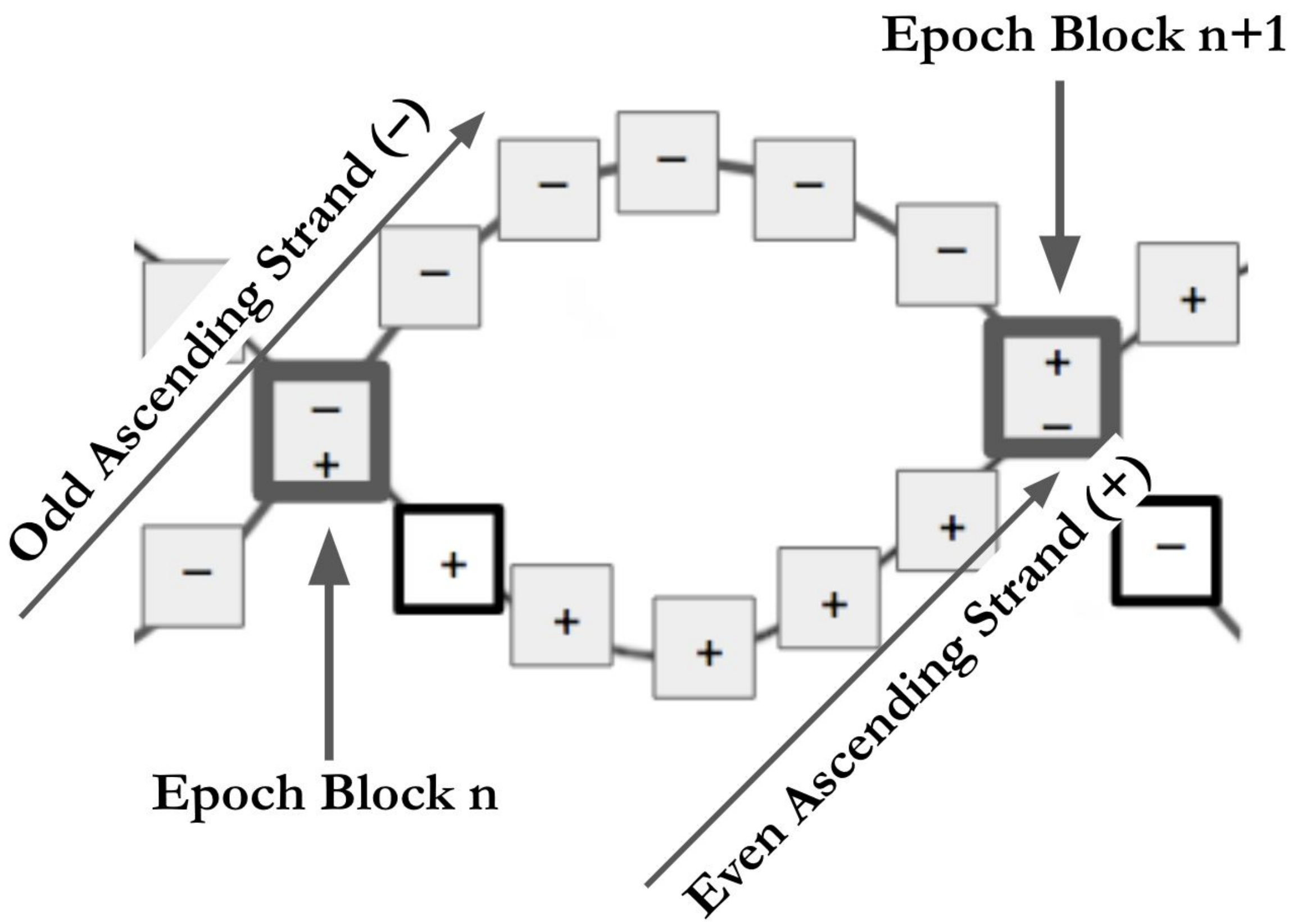

Each new Epoch Block is *proposed by the majority node configuration of the Ascending Strand* **(See Figure #4)** whose node topology is known and stored in the prior Cycle block and held by all nodes. During Epoch Blocks, the Ascending Strand's "*valence*," either positive, negative, or any other deterministic identifying mechanism, takes over the responsibilities of the descending strand's, setting the descending strand's node configuration to be set to reshuffle during said Epoch Block.

**Figure #4**

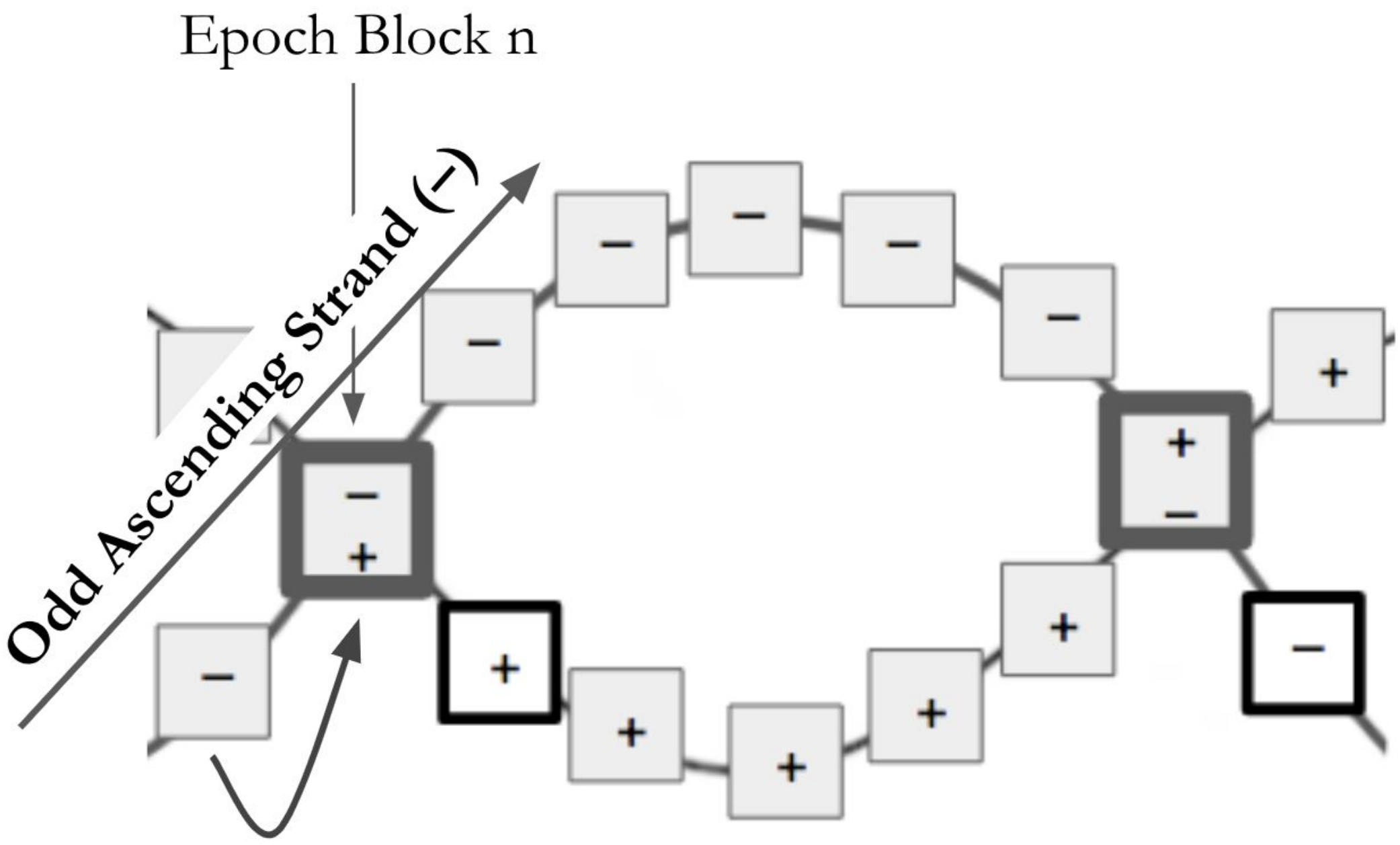

**Epoch Block n Proposed By Odd Ascending Strand (–) Majority Node Configuration**

The descending “submissive strand” will always proceed to sign a proposed Epoch Block by the ascending “dominant strand” as long as it is verifiably signed by the *majority node configuration* of the ascending strand. All nodes exist in both (all) strands of the Unitychain structure, albeit in different logical positions and also possess all blocks of each strand. Therefore, nodes are able to trivally compute if any proposed Epoch Block is correctly signed or not by both majority node configurations of the counterparting strands [5]. It is precisely these global cryptographic majority signatures that drive the protocol forward.

Once the Epoch block is cryptographically signed by both majority nodes of the intersecting Unitychain strands, the descending strand of a set begins to reshuffle its node membership, letting more nodes join the system up to a certain threshold (tunable system parameter), recreating new shared secret key pairs by performing their randomized Distributed Key Generation procedures, which should be secured to sufficiently reshuffles node membership into a newly randomized deterministic topology. Frequent and flowing randomized network membership is a prerequisite for ongoing protocol security for all chains that use Classical BFT Consensus Protocols **(See Figure #5).**

**Figure #5**

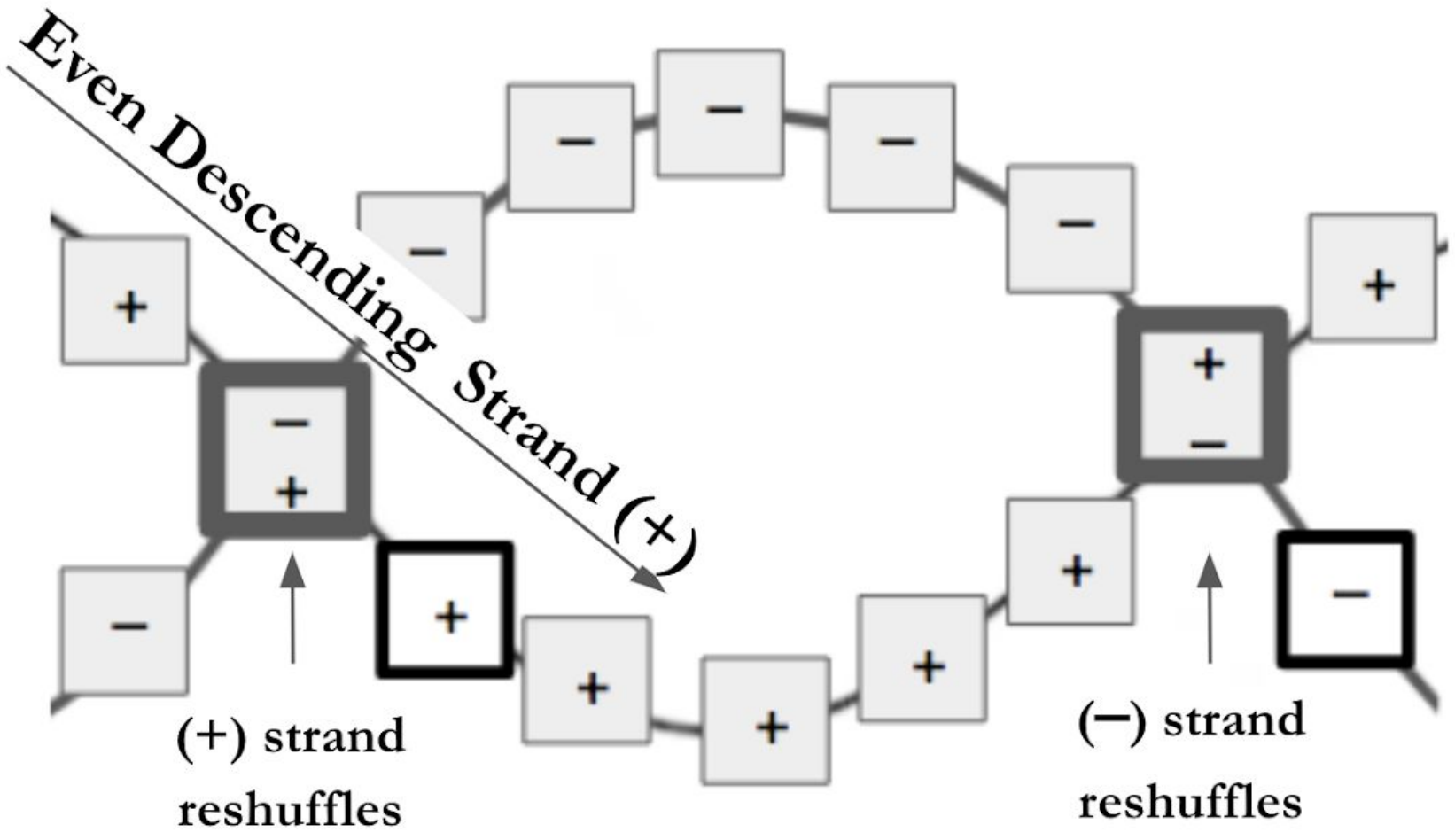

**During Epoch Blocks, the descending strand's responsibilities are assigned to the newly ascending dominant strand. The descending strand begins to reshuffle its node composition, allowing new nodes to join and leave, and creating new DKG pairings.**

Many protocols that require a Distributed Key Generation procedure to leverage threshold cryptographic signatures to drive the protocol forward *may experience down times from minutes to hours* to reshuffle node membership due to exponentially growing communication overhead and complexity based on the number of nodes involved in the key ritual [2]. Therefore, there are a multitude of other non-proof-of-work blockchains that do not reshuffle global node membership frequently, in fact, they may shuffle on the order of weekly [4] instead of every couple of minutes or hours, as is enabled by the Unitychain structure.

Since the dominant strand takes on the full network responsibilities for consensus while the other strand is reshuffling, from an end-user point of view the network throughput may be imperceptibly affected, unless the network is at peak capacity. At the very least, the end-user should *never* experience a full liveness interruption where the network is fully down and unable to process transactions during membership shuffling and joining procedures.

Non-interruption of the entire network is important for mission-critical systems that must have extremely high liveness guarantees.

Networks that do not shuffle nodes frequently inevitably suffer from less overall security due to a more static network membership, which over time is more susceptible to adversaries colluding, denial of services attacks, and other forms of coordinated corruption. With a Unitychain structure, since we are able to continuously reshuffle node membership without suffering global downtime, we are able to do this reshuffling more often, providing *substantially more safety* compared to other protocols that don't reshuffle node membership as frequently.

After the "submissive strand" of the set is complete with its reshuffling and new distributed key generation procedures, this strand proposes a new block, which contains the new network topologies and node identity data, et al., to the current dominant strand's majority node configuration.

This signals a *readiness* to continue once more with splitting the responsibilities of the network to multiple globally known node configurations, for example, dividing the database key-range or compute responsibilities into even and odd account numbers, thereby, forming newly individuated strands that are cryptographically proven to have been signed by the correct majority nodes of the prior block. At least 51% of the nodes from the prior known network configuration *must again be present in the next proposed block*, albeit in different logical positions per strand, in order to guarantee the correct cohort of nodes are included when the protocol is driven forward block by block.

The newly proposed block by the submissive strand is called an "Epoch Genesis Block" **(See Figure #6)** and also requires *both majority nodes* from the Epoch Block dominant node configuration and the node configuration of the submissive strand to cryptographically majority sign in order to be valid.

**Figure #6**

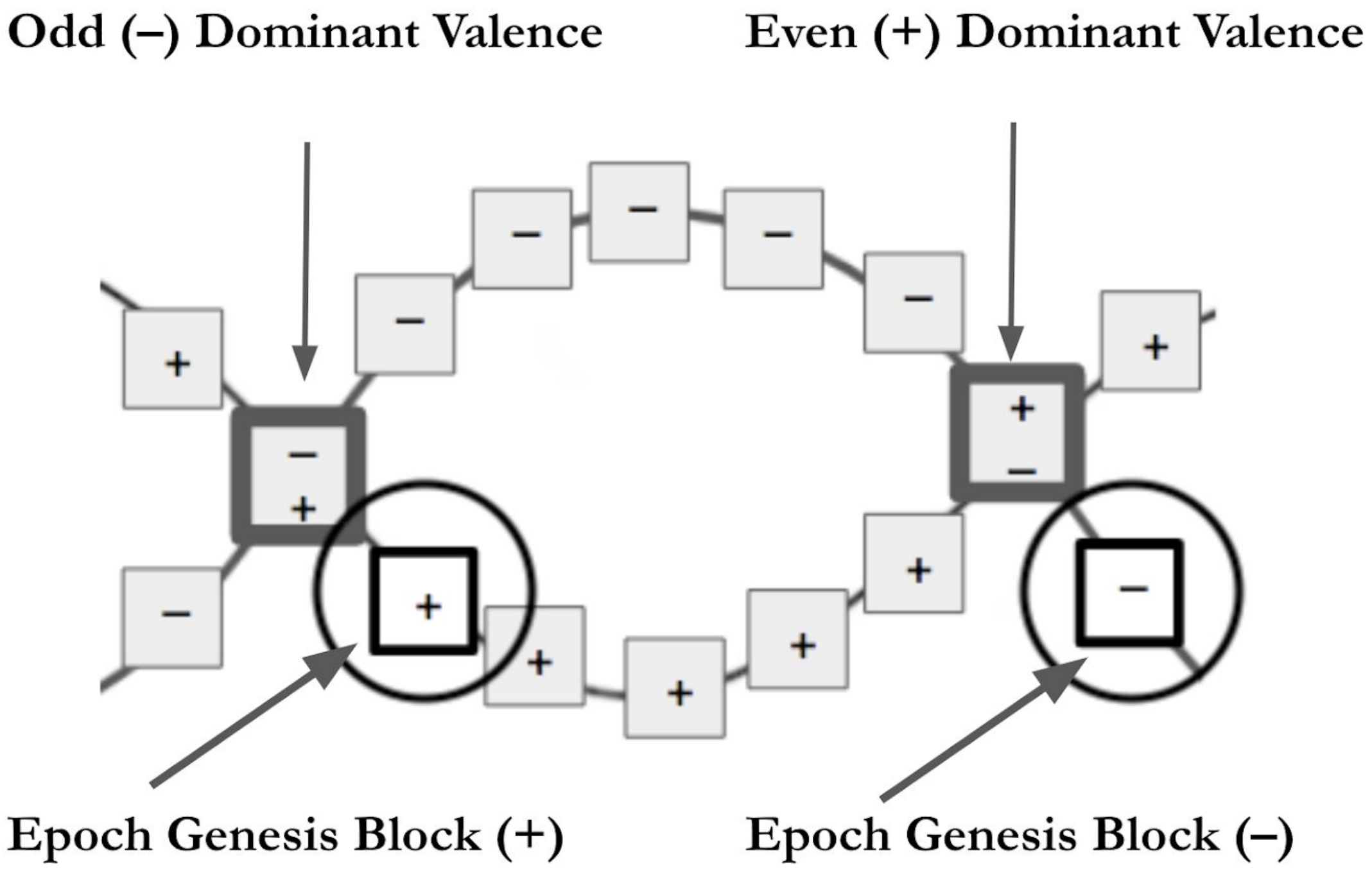

Epoch Genesis Blocks contain the newly formed network configuration data as well as the proposed bifurcation of network responsibilities between the multiple node topologies. Upon majority consensus by both strands of the set, the Unitychain once again diverges into 2 (or more) distinct strands, each responsible for managing different parts of the global database or other network services.

In addition to greater security, the Unitychain design enables nodes to more maximally leverage their full compute potential, and as a natural corollary, increase the network's *overall* scalability. Nodes may process requests and provide network services in multiple "Parallel Universes," which in essence means the physical node exists in multiple strands in different randomized topological positions untethered by its actual geo-location, with perhaps the only rule being that nodes may not play an active leadership role in both Parallel Universes in the same cycle.

The Unitychain structure more efficiently utilizes a node's full compute potential by allowing it to process transactions from a multitude of deterministic routing and processing requests. A Unitychain enables nodes to provide more useful computation contributions to the network, thereby increasing overall network scalability on an individual *node by node basis*, without any need to adjust the underlying hardware of existing validating nodes. This structure should be able to increase the throughput of virtually any blockchain that utilizes Classical BFT Consensus algorithms [8].

## Some Clear Advantages of the Unitychain Structure:

The Unitychain *multi-helix* blockchain structure enables nodes to exist in multiple network configurations allowing them to participate in various logical partitions of the global network. Perhaps the most unique aspect of our design is that during "Epoch Blocks" where multiple "strands" of a Unitychain converge into a nexus point, one node network configuration contained in one strand may deterministically take over the responsibilities of its counterpart, while the network configuration of the other strand is able to begin its reshuffling procedures, letting new nodes join and leave [9], and performing new "distributed key generation" pairings in small groups and/or as a whole. This can all be coordinated and validated through majority cryptographic signatures on Epoch Blocks of intersecting strands.

The frequent reshuffling of the node network distribution is necessary to generate more randomness and unpredictability, i.e. the more the node composition of the network is shuffled, the more difficult it is for an adversary to collude or stage an attack. Another wonderful hallmark of a Unitychain design is that the end user may not experience any down time during this reshuffling experience, since during Epoch Blocks one strand takes on the responsibilities of the other. Most other protocols that utilize DKG and threshold cryptography may require a temporary downtime ranging from minutes to hours, therefore, they do not reshuffle node membership as often, sometimes only weekly, and as a result are that much less secure. Furthermore, some protocols must endure setup times of almost a full day in order to bootstrap their network. Additionally, these protocols also tend to have a difficult time reconfiguring node membership without affecting network latency and performance [10].

## An Overview of a Unitychain in Action:

At the beginning of a Unitychain is a singular "Origin Block," which contains the original node identity membership of all nodes at the beginning of the network, as well as any other relevant data required for the protocol to initiate. The network may begin at this stage, with the single node configuration covering the responsibilities of the entire network and database.

All nodes will thereafter begin a sub-routine to organize the existing membership into one or more additional network configurations. Nodes must come to agreement of this and achieve a majority vote on the additional network configurations as well as the bifurcation of network responsibilities, i.e. which new network configuration will randomly and deterministically cover which key-range of the database or arbitrary compute partitions. The majority aggregate signature, which produces a verifiable random number when transformed into an alpha-numeric string, can be used as a Verifiable Random Function (*VRF*) [11] to assign different nodes to different positions of the protocol in each of the splitting strands of the Unitychain structure.

Upon agreement of this, by following the majority signature, the Origin Block splits into 2 (or more) individual strands, with each strand containing different network configurations of the totality of nodes that are responsible for managing consensus and other network responsibilities for the various agreed upon parts of the database or compute services.

**As depicted in Figure 7 below**, when strands diverge from an Epoch Block after their shuffling and DKG procedures are complete, the Epoch Genesis Block is first proposed (Step 1) and majority signed by the Epoch Block's submissive strand's nodes (Step 2). Next, the Epoch Genesis Block is approved by the Epoch Block's dominant strand's node configuration (Step 3) with a verifiable majority signature. Upon receiving these valid aggregate signatures, the Epoch Genesis Block's data becomes the live and active network topology.

In rapid succession, in a similar regard, the first Cycle Block of the Ascending Strand is proposed and signed by the Epoch Block's dominant node majority (Step 4) and then also signed by the node majority contained in the newly agreed upon Epoch Genesis Block (Step 5). These majority threshold signatures can be aggregated to create a final signature, that may be used as a *VRF* when computed with the node data held in the Epoch Block, in order to create a randomized permutation of the network membership that will be the new topology for this first Cycle Block of the Ascending Strand. This is how nodes, who may exist in "Parallel Universes" in various logical configurations, are able to *non-interactively* [12] have the correct view of each other's network topology, status, state, and progress *at all times.*

At this stage, the 2 or more strands of the Unitychain have diverged and each subsequent Cycle Block of each strand is majority signed by the node configuration of the prior block. This majority signature is then utilized as a *VRF* to modulate the node membership topology data for each successive block (Step 6a and 6b).

The next Epoch Block is *always* proposed by the more recently reshuffled strand (Step 7). Epoch Blocks assign the different node configurations different responsibilities, for example, one cohort of nodes shall be assigned to reshuffle their membership, while the other strand will take up the full range of responsibilities of its reshuffling counterpart. Upon seeing a *valid* Epoch Block proposal by the Ascending strand, the Descending strand will sign the proposed Epoch Block. Once the proposed Epoch Block is majority signed by both counteracting strands of the Unitychain structure, this new Epoch Block's node configuration and assignment of responsibilities becomes active (Step 8).

**Figure 7. The Unitychain Cryptographic Signature Flow of Diverging and Converging Strands**

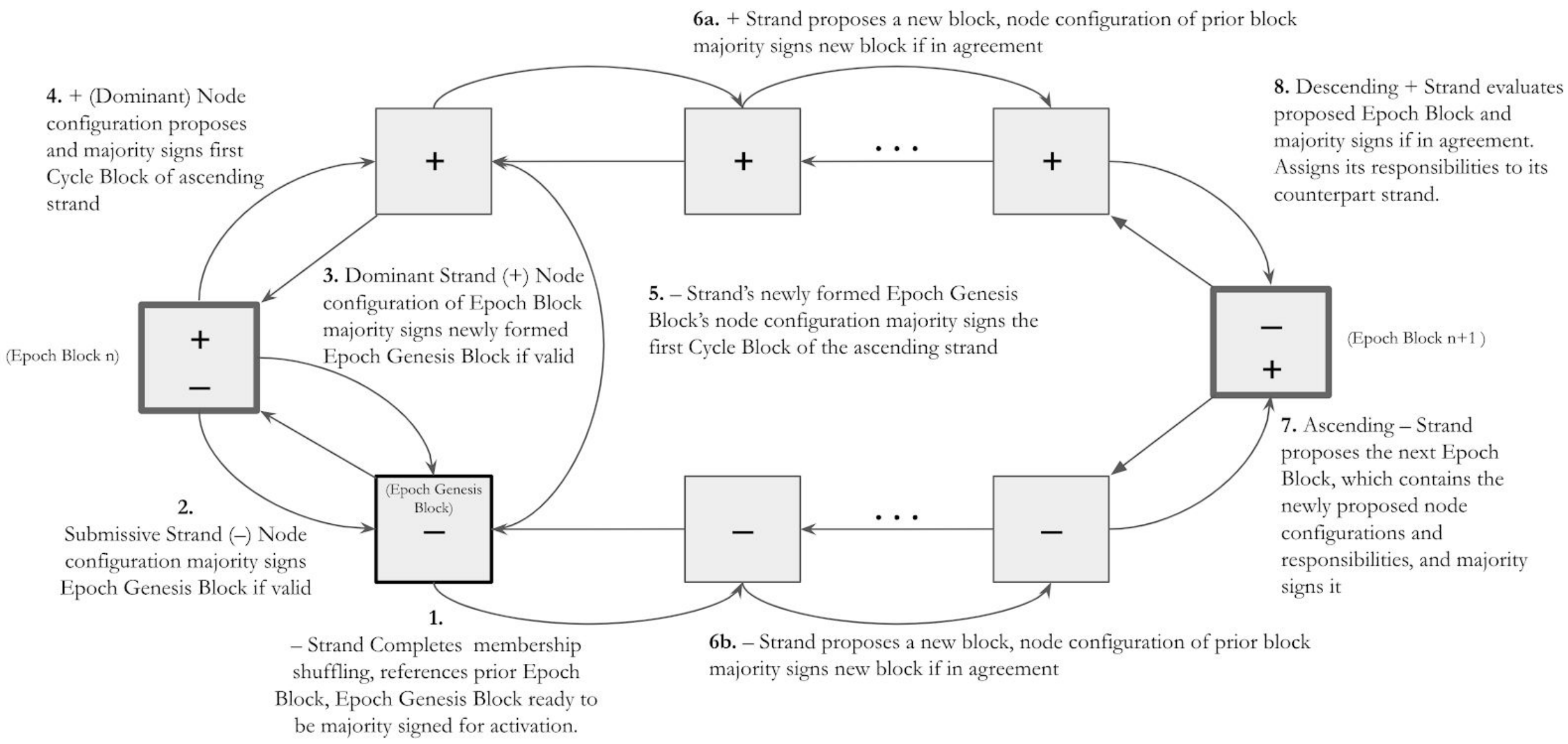

What's interesting to note here is that all nodes simultaneously exist in both (all) strands of the Unitychain structure. Additionally,  all nodes may hold the total global state *(this may be mandatory)* which we suspect will be under 200GB for a long time. At a very minimum each node may have the network topologies of potentially a large multitude of strands saved immutably and kept up-to-date. In this way, nodes always have full visibility to the continuous majority decisions of each strand of the Unitychain structure.

Verifying cryptographic signatures that prove that the correct majority of the correct network node configurations have signed *is trivial* since full knowledge of the network node composition of all strands are known to all nodes. Afterall, as was just pointed out, all nodes may in fact be actively involved in different node configurations within all strands!

## Conclusion:

This process continues back and forth with the prior ascending "dominant strand" becoming the descending "submissive strand" for the next Epoch, reshuffling and repeating these steps in a cyclical fashion to deterministically randomize node composition, thereby enabling frequent network shuffling to introduce more unpredictable entropy to the system and thus greater security for the global interweaving organism. This makes it more difficult for adversaries to collude or predict outcomes [13], while also never subjecting end-users of the network to experiencing a full downtime while nodes recalibrate their cryptographic key pairings.

This intersection of two or more block-and-chain strands to form a Unitychain brings forth remarkable *scalability* and *usability* qualities that according to our knowledge are not possible in traditional blockchain designs. The bifurcation of node responsibilities into strands enables greater parallel processing, thereby increasing the overall throughput of any network that deploys our technique.

∞